\documentclass[useAMS,usenatbib]{mn2e}  
\usepackage{graphicx,amsmath,amssymb,afterpage,color}
\def\aj{AJ}%
\def\apj{ApJ}%
\def\apjl{ApJ}%
\def\aap{A\&A}%
\def\aaps{A\&AS}%
\def\mnras{MNRAS}%
\def\pasj{PASJ}%
\def\rmxaa{Rev. Mexicana Astron. Astrofis.}%
\def\nat{Nature}%
\newcommand \radm{rad~m$^{-2}$}
\newcommand \sgr{Sgr~A$^\star$}

\begin{document}
   \title[]{Radio polarimetry of Galactic centre pulsars}

   \author[Schnitzeler et al.]
           {D.H.F.M. Schnitzeler$^{1}$\thanks{Schnitzeler@mpifr-bonn.mpg.de}, 
            R.P. Eatough$^{1}$,
            K. Ferri\`ere$^{2}$, 
            M. Kramer$^{1}$,
            K.J. Lee$^{3}$,
            \newauthor
            A. Noutsos$^{1}$,
            R.M. Shannon$^{4,5}$\\
          $^1$ Max Planck Institut f\"ur Radioastronomie, D-53121 Bonn, Germany\\
          $^2$ IRAP, Universit\'e de Toulouse, CNRS, BP 44346, 31028 Toulouse Cedex 4, France\\
          $^3$ Kavli Institute for Astronomy and Astrophysics, Peking University, Beijing 100871, People's Republic of China\\
          $^4$ Australia Telescope National Facility, CSIRO Astronomy and Space Science, Marsfield, NSW 2122, Australia\\
	 $^5$ International Centre for Radio Astronomy Research (ICRAR), Curtin University,  Bentley, WA 6102, Australia\\ 
          }

   \date{Accepted 2016 April 08. Received 2016 April 07; in original form 2016 January 27.}

   \pagerange{}\pubyear{}

   \maketitle

\begin{abstract}
To study the strength and structure of the magnetic field in the Galactic centre (GC) we measured Faraday rotation of the radio emission of pulsars which are seen towards the GC.
Three of these pulsars have the largest rotation measures (RMs) observed in any Galactic object with the exception of \sgr.
Their large dispersion measures, RMs and the large RM variation between these pulsars and other known objects in the GC implies that the pulsars lie in the GC and are not merely seen in projection towards the GC.
The large RMs of these pulsars indicate large line-of-sight magnetic field components between $\sim\,16-33\,\mu$G; combined with recent model predictions for the strength of the magnetic field in the GC this implies that the large-scale magnetic field has a very small inclination angle with respect to the plane of the sky ($\sim\,12\degr$).
Foreground objects like the Radio Arc or possibly an ablated, ionized halo around the molecular cloud G0.11-0.11 could contribute to the large RMs of two of the pulsars.
If these pulsars lie behind the Radio Arc or G0.11-0.11 then this proves that low-scattering corridors with lengths $\gtrsim\,100\,\mathrm{pc}$ must exist in the GC.
This also suggests that future, sensitive observations will be able to detect additional pulsars in the GC.
Finally, we show that the GC component in our most accurate electron density model oversimplifies structure in the GC.
\end{abstract}

%

\section{Introduction}\label{introduction.sec}
The distribution of ionized gas and the strength and structure of the magnetic field in the GC\footnote{
Throughout this paper we will refer to the inner $\sim\,150\,$pc region centred on \sgr\ as `the GC', which corresponds to the region with strongly enhanced electron densities in NE2001 \citep{cordes2002}.
} have been probed using diffuse synchrotron emission (e.g., \citealt{yusef-zadeh1984}, \citealt{tsuboi1986}, and \citealt{sofue1987}) and extragalactic radio sources \citep{roy2008}. These early measurements found evidence for strong magnetic fields, dense gas, and a new phenomenon: non-thermal radio filaments (NTFs), which are thin strands or sheets of synchrotron-emitting plasma. Over the last few decades many new insights have shed light on this complex environment, but important questions regarding the strength and structure of the magnetic field that pervades the GC remain unanswered (for a critical review see \citealt{ferriere2011}).

We study the properties of the magnetic field in the GC by observing all known radio pulsars within $\approx$ 20\arcmin\ of \sgr. 
First we establish that these pulsars lie in the GC and are not merely seen towards it, then we investigate the implications for scattering in the GC.
Radio pulsars have several advantages over source types that have been used previously: pulsars are perfect point sources, pulsar RMs are produced exclusively in the Galactic foreground \citep{noutsos2009}, pulsar distances can be estimated from the delay in pulse arrival times with frequency (quantified by the dispersion measure DM), and the line-of-sight (LOS) magnetic field component in the interstellar medium (ISM) can be estimated from the amount of Faraday rotation of the pulsar signal (quantified by RM).
At a wavelength $\lambda$ the amount of Faraday rotation is equal to $\chi - \chi_0\, =\, \mathrm{RM}\,\lambda^2$, where $\chi$ and $\chi_0$ are the observed and the intrinsic polarization angle of the emission, respectively. 
\begin{eqnarray}
\mathrm{RM}\, \left(\mathrm{rad~m}^{-2}\right)\ =\ 0.81 \int_\mathrm{source}^\mathrm{observer} n_\mathrm{e} B_\| \mathrm{d}l\, ,
\label{rm_definition}
\end{eqnarray}
where $n_\mathrm{e}$ is the local electron density in units of cm$^{-3}$, $B_\|$ the LOS component of the magnetic field in units of $\mu\mathrm{G}$, and the path length $l$ is measured in parsec. 
If the LOS component of the magnetic field points towards us $B_\|$ is positive, RM increases, and the plane of polarization is rotated counterclockwise.
The dispersion measure DM is defined as
\begin{eqnarray}
\mathrm{DM}\, \left(\mathrm{cm}^{-3}~\mathrm{pc}\right)\ =\ \int_\mathrm{source}^\mathrm{observer} n_\mathrm{e} \mathrm{d}l\, ,
\label{dm_definition}
\end{eqnarray}
and the emission measure $\mathrm{EM}\, \left(\mathrm{cm}^{-6}~\mathrm{pc}\right)\ =\ \int_\mathrm{source}^\mathrm{observer} n_\mathrm{e}^2 \mathrm{d}l$.
The pulsars in our sample have been discovered by \cite{johnston2006} and \cite{deneva2009} and include also the recently discovered magnetar PSR J1745-2900 which is known to lie close to \sgr (\citealt{kennea2013}, \citealt{mori2013}, \citealt{eatough2013}, \citealt{shannon2013}).

This paper is structured as follows: in section~\ref{observations.sec} we describe the observations and their calibration, and in section~\ref{discussion.sec} we interpret our results in terms of the structure of the magnetized ISM in the GC. In section~\ref{summary.sec} we summarise our results. 
We will adopt a distance of 8.3 kpc to the GC \citep{reid2014}; at this distance 1 arcminute on the sky corresponds to a distance of 2.4 parsec.

\begin{figure*}
\caption{
Pulse profiles of the target pulsars, showing Stokes $I$ (black), $Q$ (red) and $U$ (green) after correcting for Faraday rotation using the RMs from table~\ref{pulsars.tab}, and $V$ (blue). The Stokes $I$ profile also shows 1-sigma errors. 
}
  \begin{minipage}[b]{0.42\linewidth}
    \resizebox{\hsize}{!}{\includegraphics[width=\linewidth]{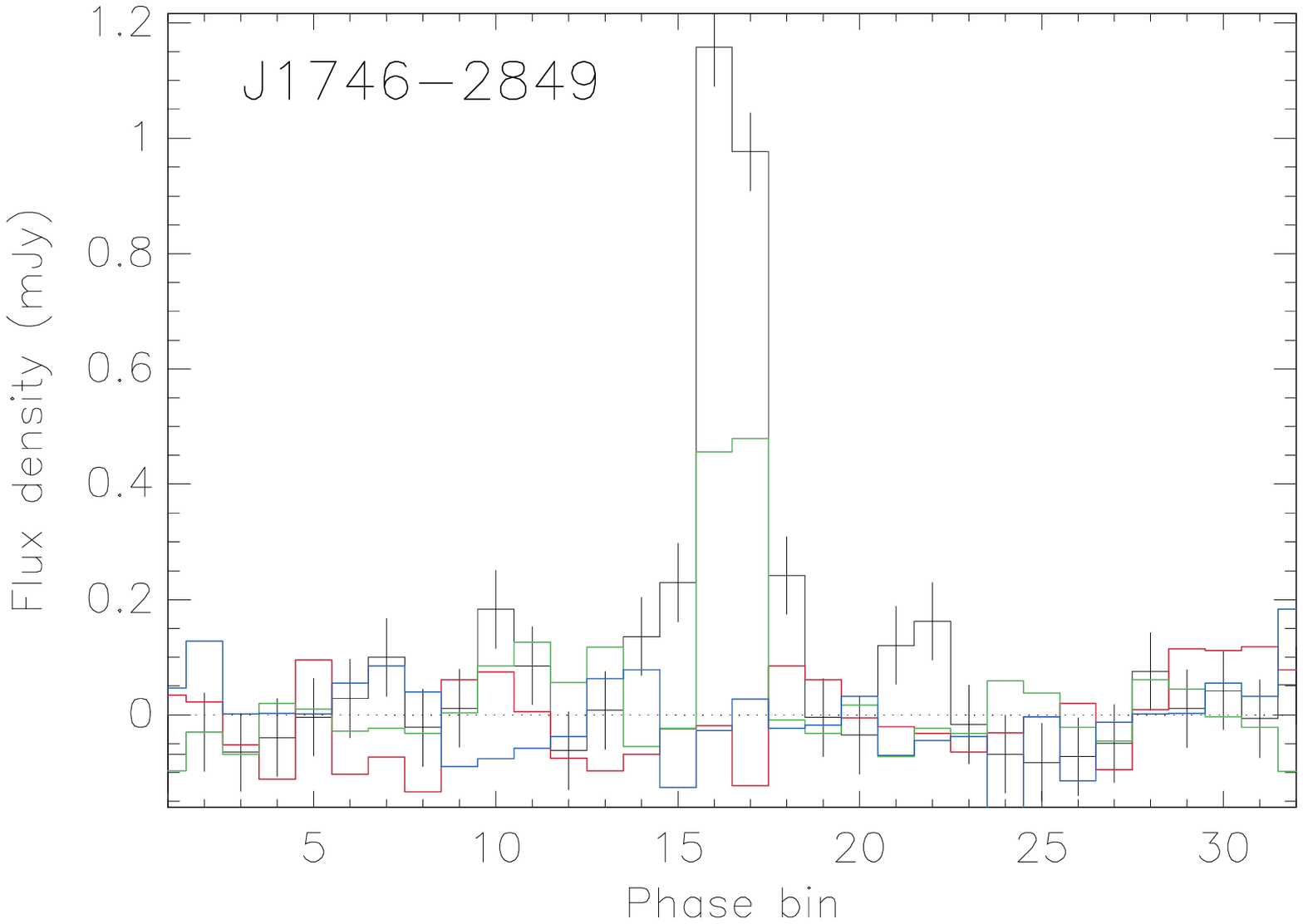}}
    \vspace{-3ex}
  \end{minipage}
  \begin{minipage}[b]{0.42\linewidth}
    \resizebox{\hsize}{!}{\includegraphics[width=\linewidth]{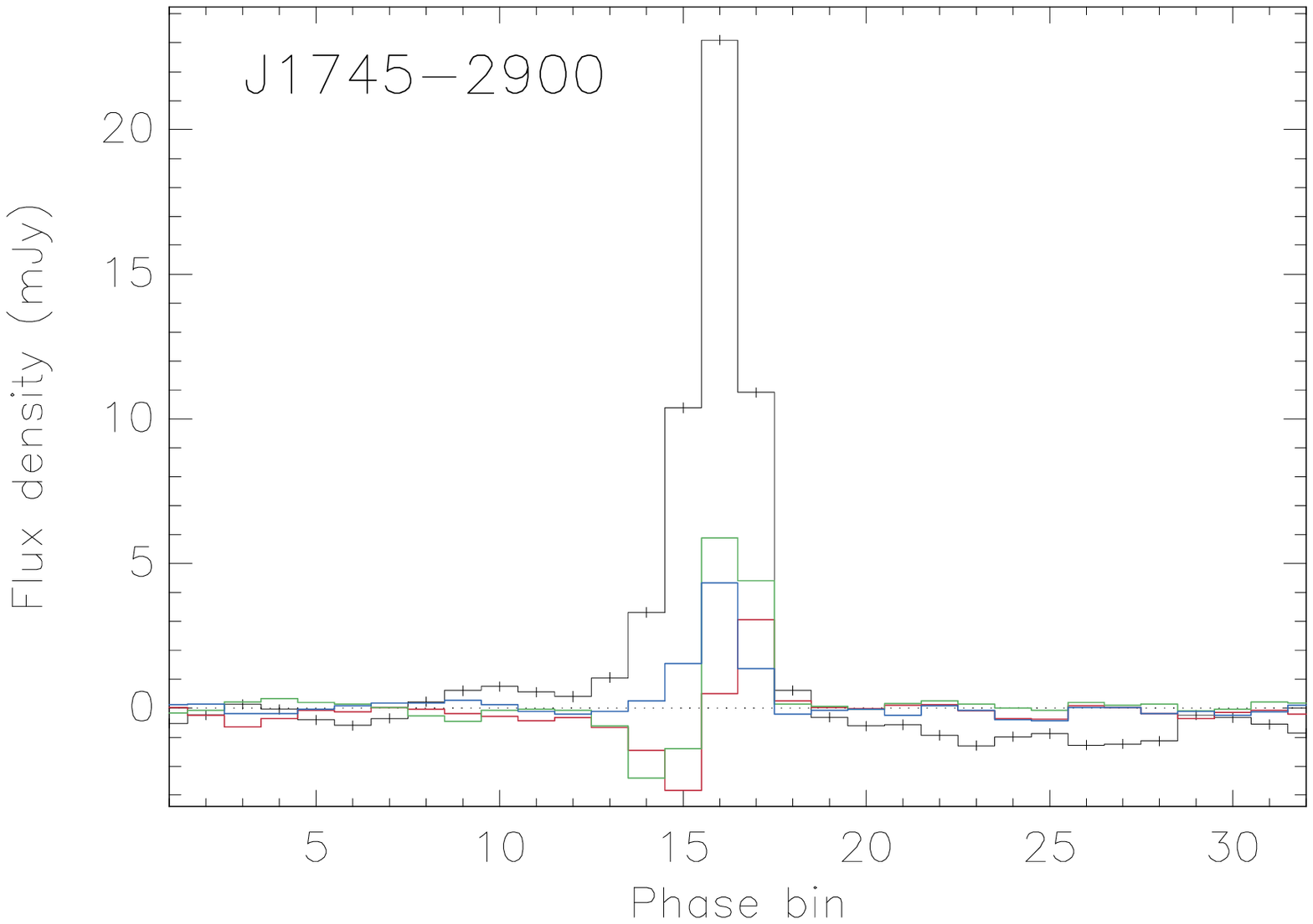}}
    \vspace{-3ex}
  \end{minipage}
  \begin{minipage}[b]{0.42\linewidth}
    \resizebox{\hsize}{!}{\includegraphics[width=\linewidth]{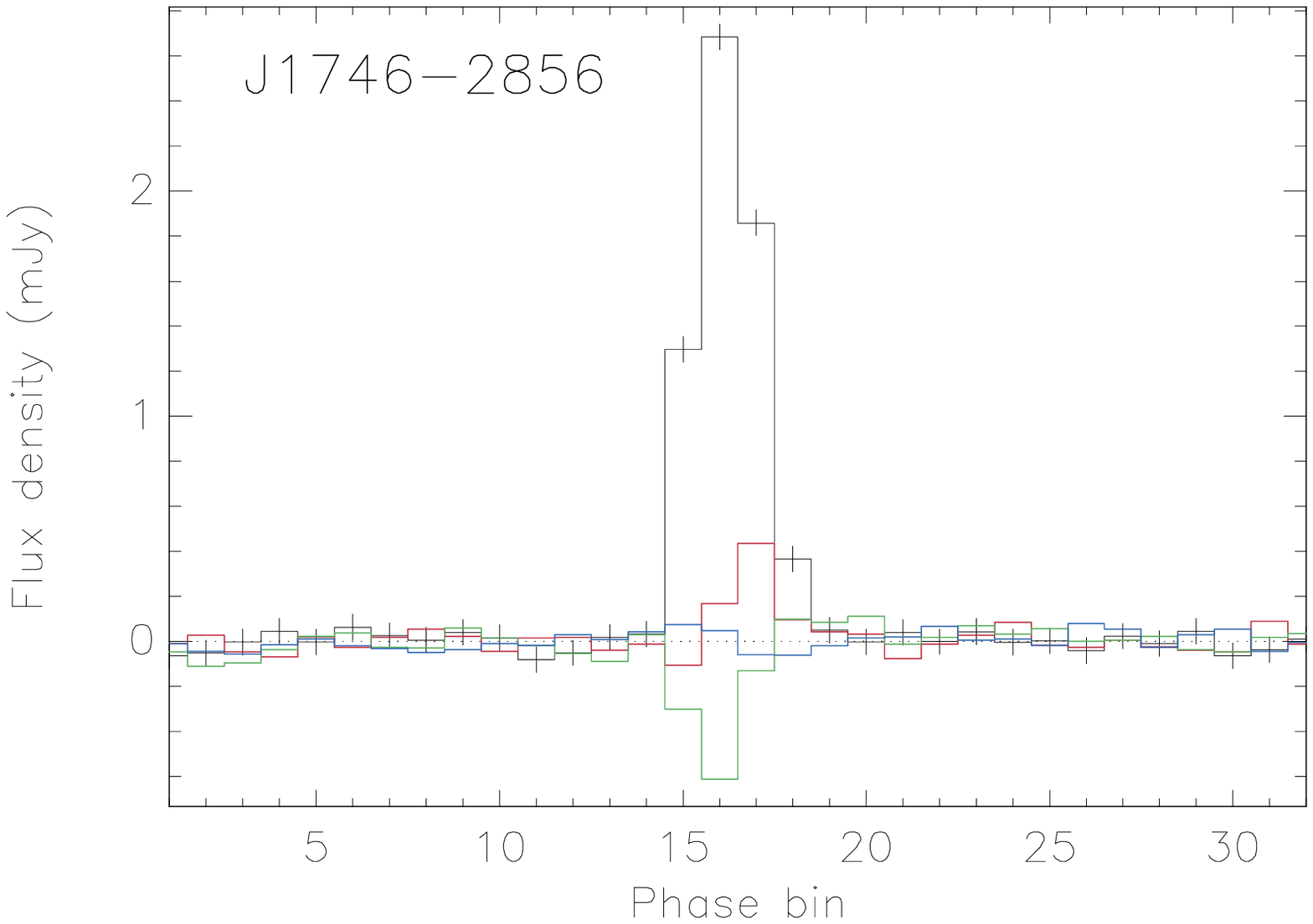}}
    \vspace{-3ex}
  \end{minipage}
  \begin{minipage}[b]{0.42\linewidth}
    \resizebox{\hsize}{!}{\includegraphics[width=\linewidth]{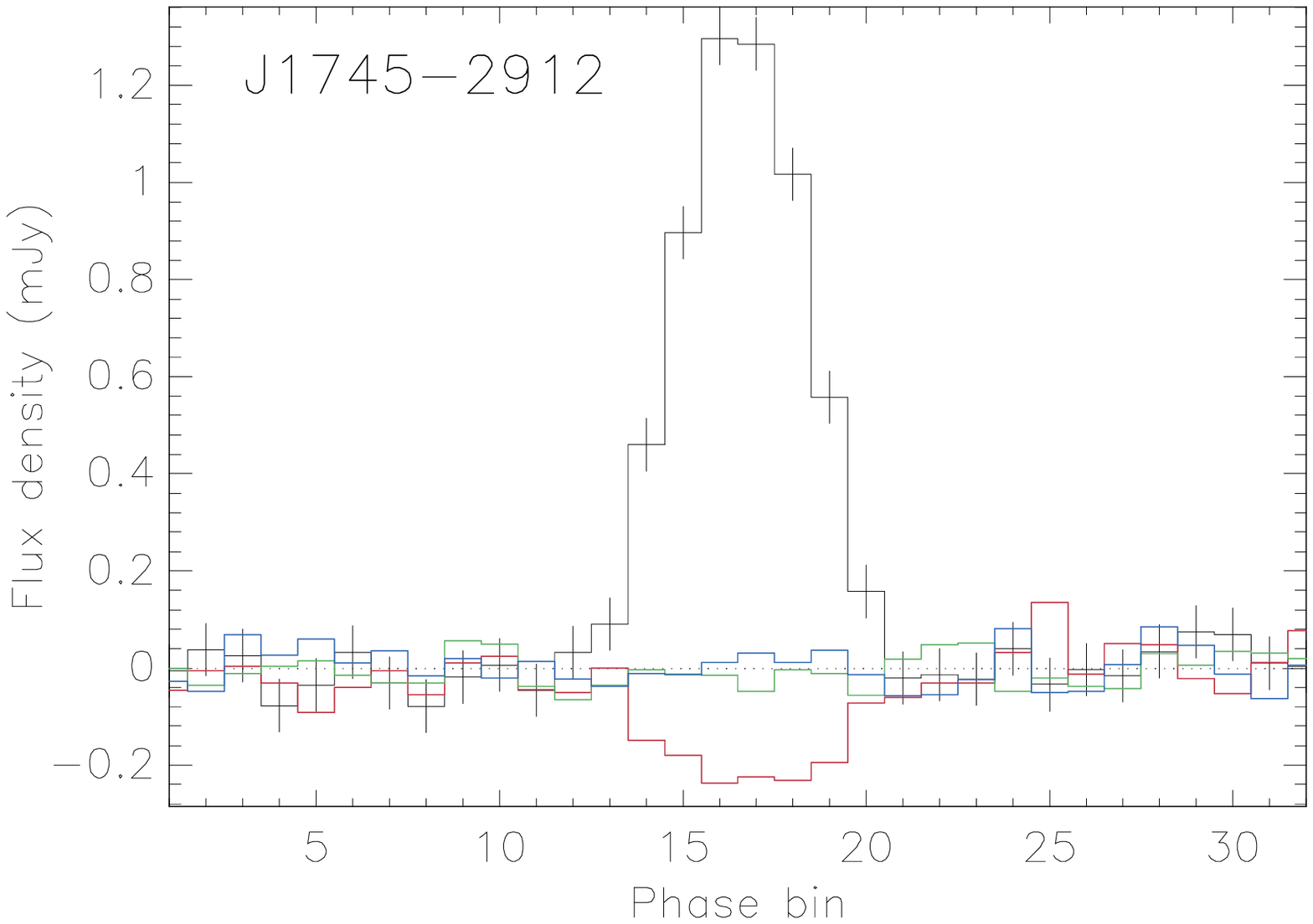}}
    \vspace{-3ex}
  \end{minipage}
\label{profiles.fig}
\end{figure*}

\begin{figure*}
\caption{Frequency spectra of the target pulsars, showing Stokes $Q$ (top panel) and $U$ (bottom panel) together with the best-fitting models and 68\% confidence intervals. 
}
  \begin{minipage}[b]{0.42\linewidth}
    \resizebox{\hsize}{!}{\includegraphics[width=\linewidth]{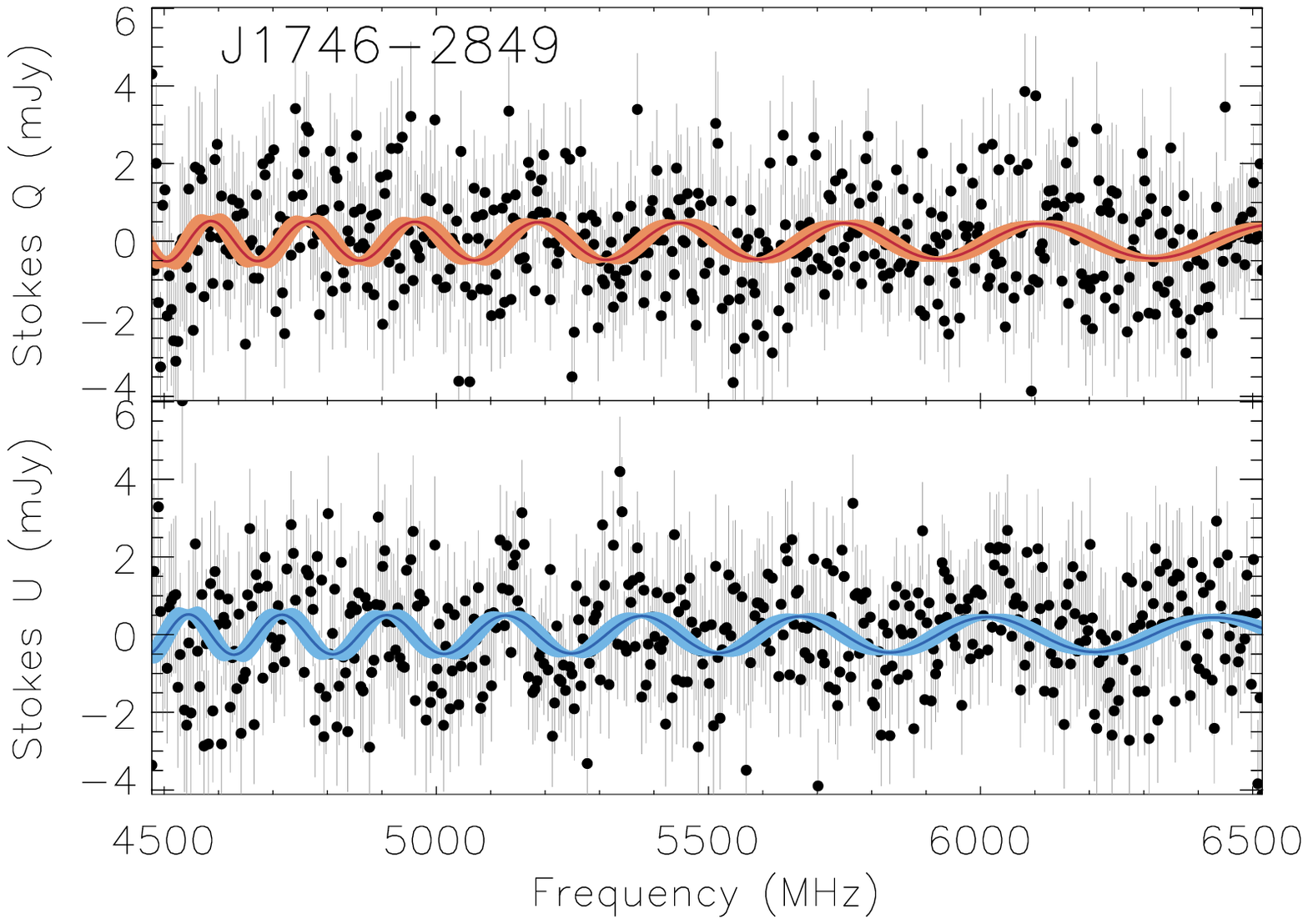}}
    \vspace{-3ex}
  \end{minipage}
  \begin{minipage}[b]{0.42\linewidth}
    \resizebox{\hsize}{!}{\includegraphics[width=\linewidth]{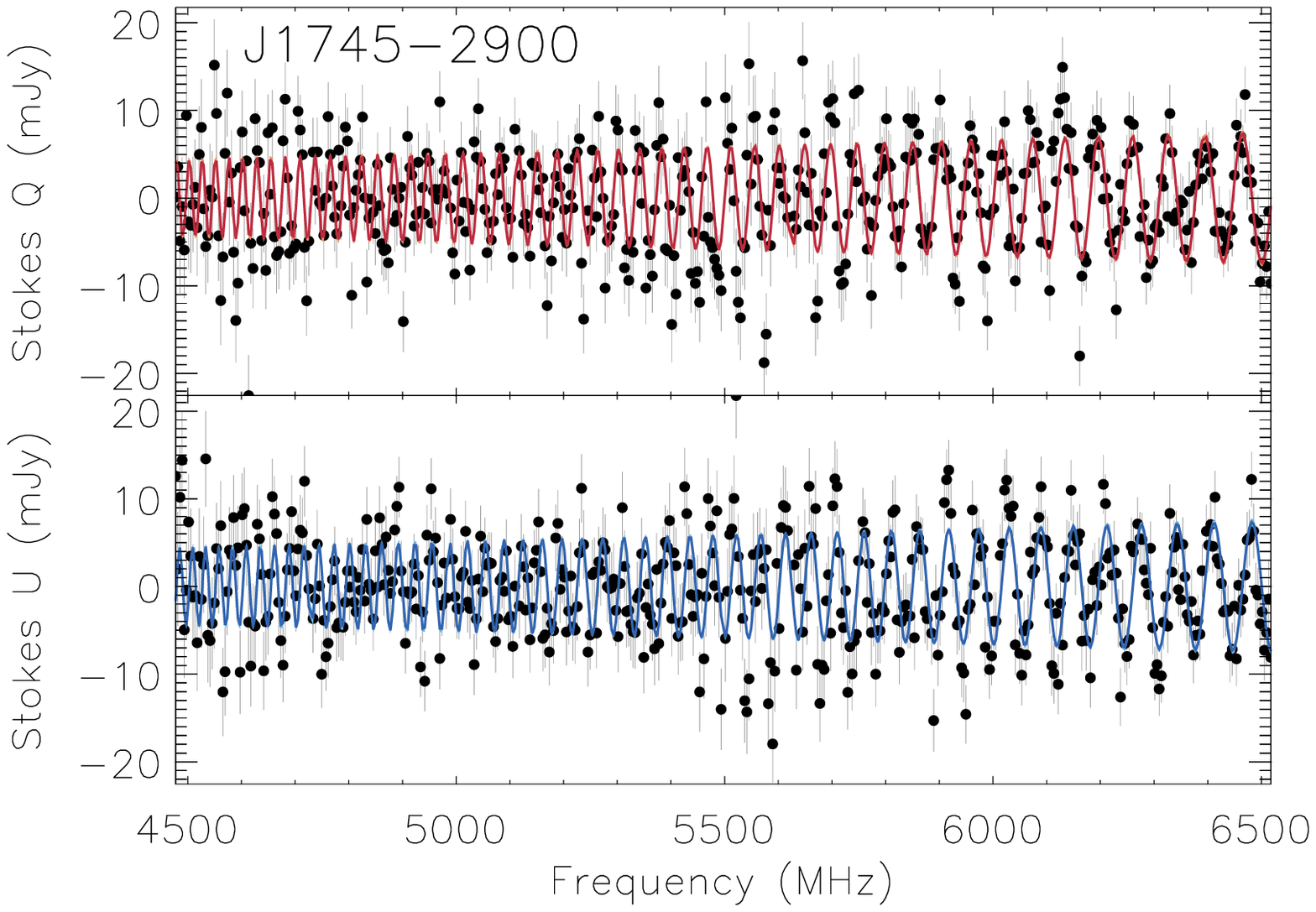}}
    \vspace{-3ex}
  \end{minipage}
  \begin{minipage}[b]{0.42\linewidth}
    \resizebox{\hsize}{!}{\includegraphics[width=\linewidth]{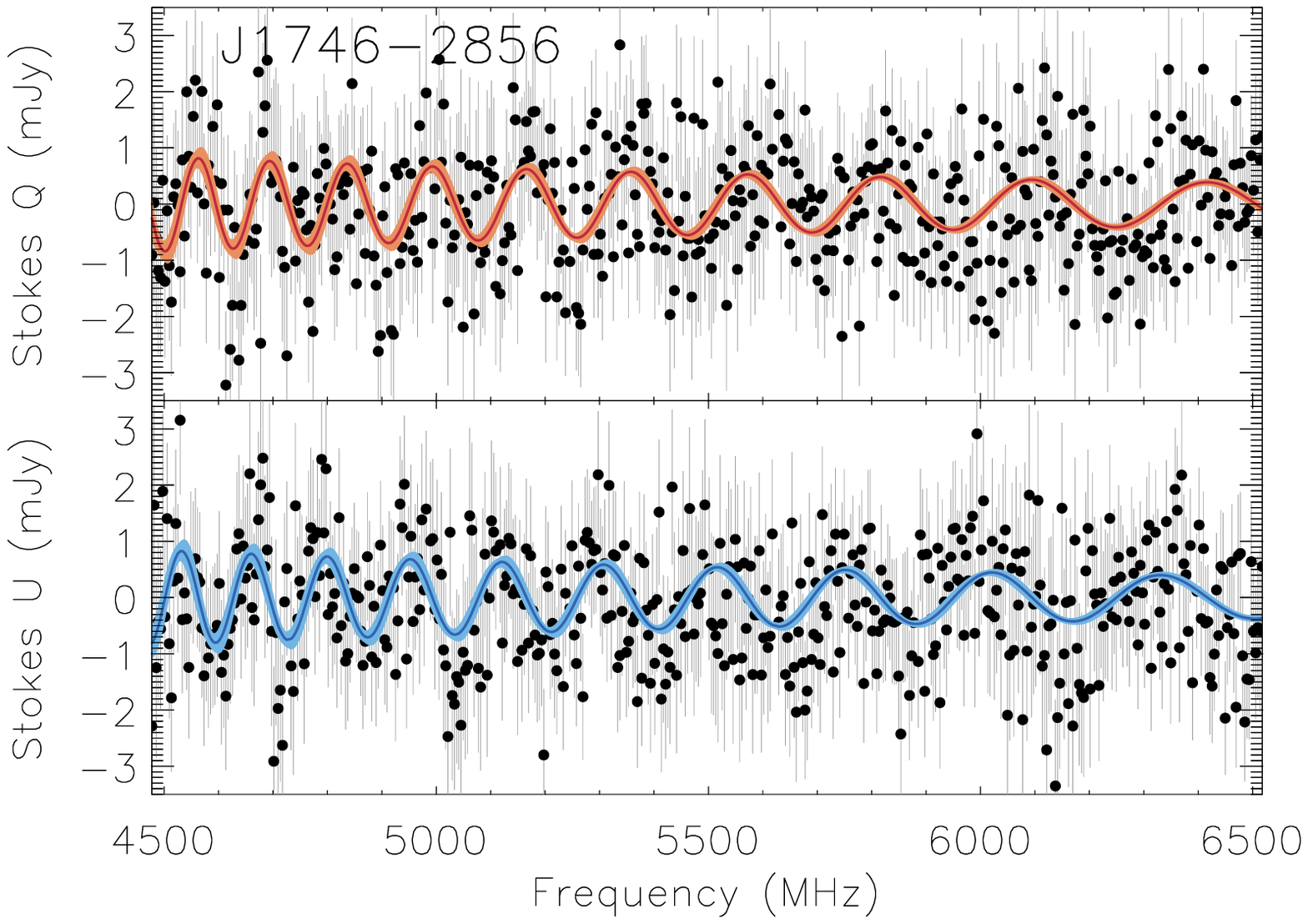}}
    \vspace{-3ex}
  \end{minipage}
  \begin{minipage}[b]{0.42\linewidth}
    \resizebox{\hsize}{!}{\includegraphics[width=\linewidth]{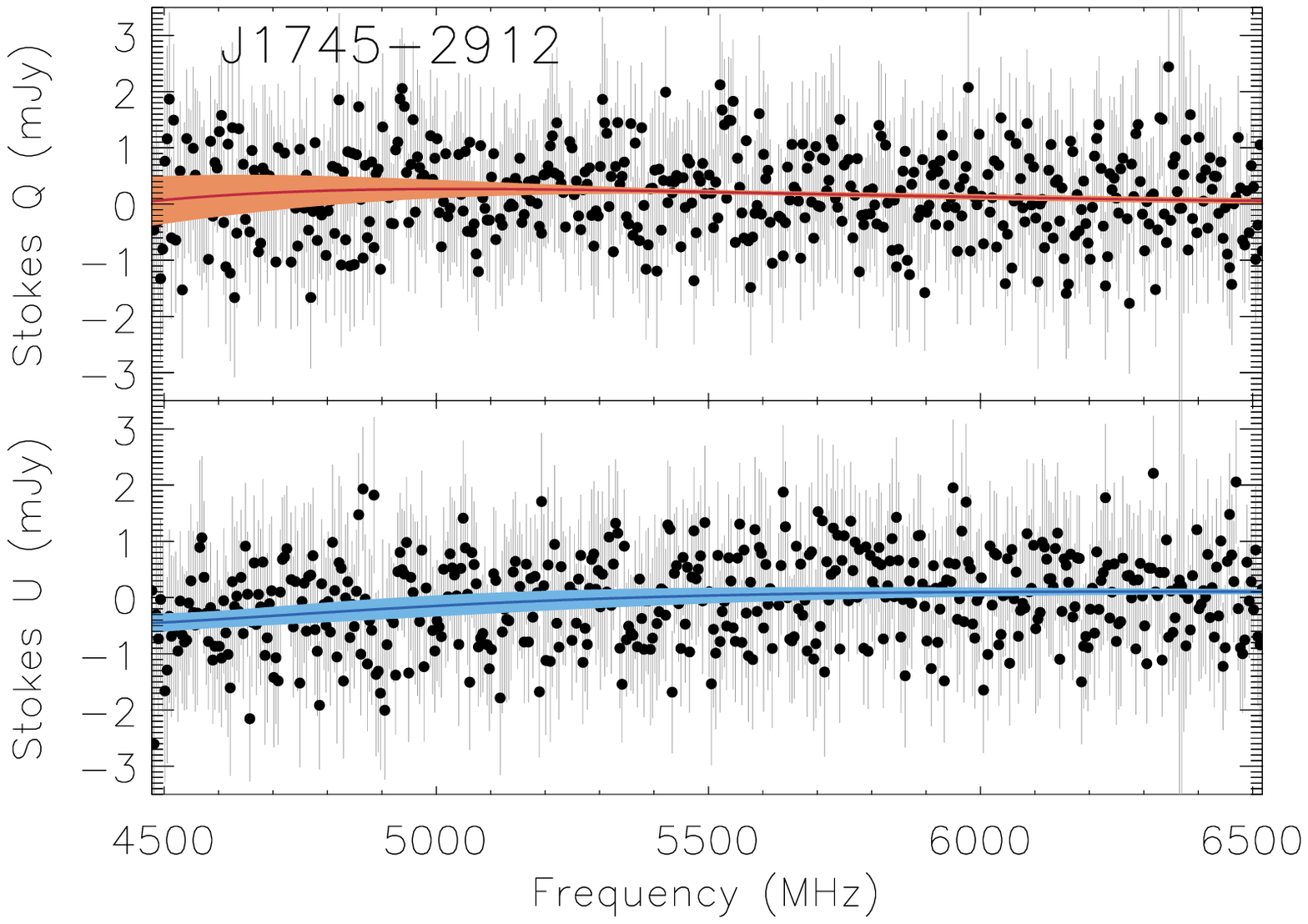}}
    \vspace{-3ex}
  \end{minipage}
\label{spectra.fig}
\end{figure*}

\begin{table*}
\centering
\caption{Properties of the pulsars that we detected: Coordinates (Equatorial and Galactic), DM, Stokes $I$ flux density of the brightest phase bin, polarized flux density and intrinsic polarization angle at 5.5 GHz ($\chi_0 = 0$\degr\ towards the west and increases counter-clockwise), flux spectral index $\alpha$, and RM. 
For comparison, the Galactic coordinates of \sgr\ are gl, gb = -0.056\degr, -0.046\degr\ \protect\citep{reid1999}.
$\chi^2_4$ lists the value of the chi-square statistic for four degrees of freedom, while ``S/N''  indicates the signal/noise level of a 1D Gaussian random variable which produces the same probability.
$^\star$: DM values were taken from \citet[`J06']{johnston2006}, \citet[`D09']{deneva2009}, and \citet[`E13']{eatough2013}. $^\dagger$: Coordinates from \citet{shannon2013}. 
}
\begin{tabular}{lcccc}
\hline
                          & J1746-2849                            & J1745-2900                             & J1746-2856                        & J1745-2912 \\
\hline
$\rmn{RA}(2000)$	  & $17^\rmn{h}46^\rmn{m}03\fs355(8)$     &  $17^\rmn{h}45^\rmn{m}40\fs16^\dagger$ & $17^\rmn{h}46^\rmn{m}49\fs856(3)$ & $17^\rmn{h}45^\rmn{m}47\fs830(8)$ \\
$\rmn{Dec.}(2000)$        & $-28\degr50\arcmin13\farcs56(23)$     & $-29\degr00\arcmin29\farcs82^\dagger$  & $-28\degr56\arcmin59\farcs23(10)$ & $-29\degr12\arcmin30\farcs77(23)$ \\
$\rmn{gl}$		  &   0.134\degr                          &  -0.056\degr                           &  0.126\degr                       & -0.212\degr \\
$\rmn{gb}$                &  -0.030\degr                          &  -0.047\degr                           & -0.233\degr                       & -0.175\degr \\
DM~(cm$^{-3}~$pc)$^\star$ &   1456 (D09)                          & 1778 (E13)                             & 1168 (J06)                        & 1130 (J06) \\
$I$~(mJy)		  &   1.16(7)                             & 23.08(17)                              & 2.68(6)                           & 1.30(5)\\
$PI_{5.5}$~(mJy)	  &   0.48(6)                             & 5.92(17)                               & 0.54(4)                           & 0.22(3)\\
$\chi_{0,5.5}$~($^\circ$) &  46.3(35)                             & 42.5(8)                                & 37.3(22)                          & -84.3(39)\\
$\alpha$ 		  & -0.48(116)                            & 1.34(28)                               & -2.18(74)                         & -3.58(140)\\
RM~(rad~m$^{-2}$)	  & 10,104(101)                           & -66,080(24)                            & 13,253(53)                        & -535(107)\\
$\chi^2_4$ (``S/N'')	  & 64 (7)                                & 779 ($>\,10$)                          & 166 ($>\,10$)                     & 54 (6)\\
\hline
\end{tabular}
\label{pulsars.tab}
\end{table*}

\section{Observations}\label{observations.sec}
The targets were observed with the Australia Telescope Compact Array (ATCA) in the 6A configuration between the 20th and the 22nd of April, and on the 6th of May 2015. These observations cover the frequency range between 4.473 and 6.525 GHz with 513 channels that are 4 MHz wide. 
PSR J1745-2900 was observed for 2.6 hours in total, other targets were observed for 5.1 hours each.
Because of telescope maintenance all six antennas were only available on the final observing run. On the first three observing runs only 4, 5, and 4 antennas, respectively, were available.
We calibrated the data using the $\textsc{miriad}$ software package \citep{sault1995}: PKS 1934-638 was used to determine the bandpass and absolute flux scale, and PKS 1814-254 (9\degr\ from the targets) was observed approximately every 18 minutes to calibrate the complex gains and polarization leakages. We then transferred the calibration solutions to the target pulsars, and flagged radio frequency interference. PKS 1814-254 rises after the target pulsars; therefore the calibration solution from PKS 1814-254 was copied to the first observation of each target without extrapolation over time. 
In our analysis we excluded the two shortest baselines.

Often the pulsar positions listed in the discovery papers had much larger errors than the size of the synthesized beam of our observations ($\approx 4\arcsec \times 1.5\arcsec$),
and we had to search maps of the primary beam to identify the pulsars. We removed a baseline level, calculated from off-pulse bins, before combining the real parts of the visibilities of the four observing runs using the noise variance in Stokes $V$ as weights. Noise-weighted pulse profiles are shown in fig.~\ref{profiles.fig}, and table~\ref{pulsars.tab} lists the pulsars that we detected with the ATCA, their updated coordinates, and the peak total intensity. 
Positions were determined after summing the two brightest phase bins from the data collected in May 2015. 

The pulsars listed in Table~\ref{pulsars.tab} are bright enough that we could detect them in our most sensitive observing run. 
We were not able to detect PSR J1746-2850, even after concatenating visibilities - across all possible pulse phase bin combinations - from our two most sensitive observing runs\footnote{The ATCA depends on pulsar ephemerides to generate pulse profiles; it is not possible to record baseband or pulsar search data. Although we used the most recent available ephemerides, the non-detection of PSR J1746-2850 could also indicate that the ephemeris for this pulsar was no longer accurate enough to fold the data correctly.}, giving a six-sigma limit on the peak flux density of 0.49 mJy.
This non-detection implies that the flux density of this source has decreased by at least a factor of 30 since it was discovered by Deneva et al. 
Radio magnetars are known to exhibit large variations in flux density (e.g., \citealt{lazaridis2008}); if PSR J1746-2850 shares characteristics with radio magnetars, as was proposed by Deneva et al., then this could explain our non-detection.
Monitoring observations might be able to detect this source in case it brightens again in the future.
We excluded PSR J1745-2910 from our analysis because of its poorly constrained position and pulse period.

We determined the polarization properties of the pulsars using a new maximum-likelihood-based method that we developed, which we describe in a forthcoming paper (Schnitzeler et al., in prep.). In our method the pulsar is characterized by its polarized flux density and intrinsic polarization angle at 5.5 GHz, its flux spectral index $\alpha$ ($S_\nu \propto \nu^\alpha$) and RM, plus, we allow the noise variances to be off by a scale factor $\eta$. Assuming that the differences between the measurements of Stokes $Q$ and $U$ and the model are described by Gaussian random variables and that the $N_\mathrm{ch}$ frequency channels are independent, we maximize the log likelihood over the parameter space. We search for RMs out to $\pm\, 4\times10^5$ rad~m$^{-2}$, the equivalent of the `maximum RM' from \cite{brentjens2005} in RM synthesis\footnote{Introducing $\delta\lambda^2$ for the channel width in units of wavelength squared, the `maximum RM' satisfies the equation RM$_\mathrm{max}\delta\lambda^2\approx \sqrt{3}$, even though RM synthesis assumes that for all frequency channels RM$\delta\lambda^2\ll1$ (both expressions are from \citealt{brentjens2005}). The difference between the exact formalism by \cite{schnitzeler2015} and the formalism by \cite{brentjens2005} is small for our observing setup and the RM range that we chose.}, and $\alpha$ between -6 and 2, which covers all pulsar spectral indices that were determined by \cite{Lorimer1995} and \cite{Bates2013}. 
The difference between the maximum log likelihood and the log likelihood in the absence of any signal follows a chi square distribution \citep{wilks1938} with in our case four degrees of freedom; this enables us to quantify the detection probability of each pulsar. The 68\% confidence limits for the pulsar parameters are based on the region where the log likelihood has decreased by 0.5 (e.g., \citealt{avni1976}).

\begin{figure}
\caption{Known GC pulsars overplotted on an Effelsberg single-dish total intensity map of the GC at 10.55 GHz \protect\citep{seiradakis1989}. The positions of the Quintuplet and Arches clusters are indicated with pluses, and pulsars that we did not detect are highlighted in orange. On this scale PSR J1745-2900 and \sgr\ lie in the same direction. 
}
\resizebox{\hsize}{!}{\rotatebox{-90}{\includegraphics[width=\linewidth]{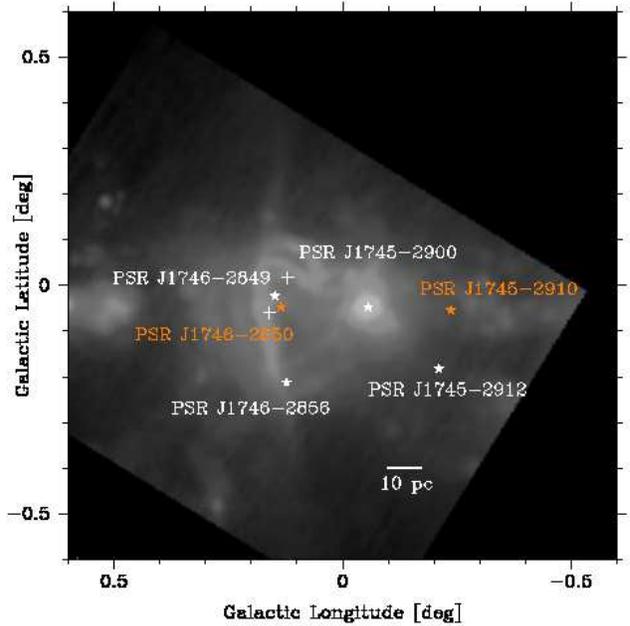}}}
\label{map.fig}
\end{figure}

Fig.~\ref{spectra.fig} shows measured and modelled frequency spectra for the target pulsars, and our maximum likelihood estimates of the polarization properties of each pulsar are listed in table~\ref{pulsars.tab}. Fig.~\ref{map.fig} shows the positions of the pulsars relative to other features in the Galactic centre.
Scintillation of PSR J1745-2900 leads to a systematic and statistically significant variation between the amplitudes of the observed and modelled spectra across the frequency band.
To increase the signal strength in our maximum likelihood estimation we summed the polarization vectors of the brightest two phase bins in the pulse profile of PSR J1746-2856 and PSR J1746-2849, while for PSR J1745-2912 we summed the brightest four phase bins.
If the intrinsic polarization angle $\chi_0$ changes across the pulse profile this leads to depolarization if the polarization vectors of different phase bins are added. To mitigate this effect we added a phase gradient to the polarization vectors of the different phase bins before summing the bins. We varied the gradient between 0\degr/bin and 180\degr/bin, and used the gradient which produced the strongest polarized signal. The largest phase gradient is 29\degr/bin.

\section{Discussion}\label{discussion.sec}
The large DMs suggest that all target pulsars lie within $\approx$ 150 pc of \sgr\ (based on the free electron density models NE2001 and NE2001thick, \citealt{cordes2002}, \citealt{schnitzeler2012}), within the Central Molecular Zone and the footprint of the Galactic Centre Lobe \citep{sofue1984}.
\cite{johnston2006} and \cite{deneva2009} noticed that the pulse broadening times of the pulsars in our sample are much shorter than predicted by these electron density models.
The absence of strong scattering does not preclude these pulsars from being in the GC: as shown by \cite{lazio1999}, \cite{bower2001}, and \cite{roy2013}, the observed scattering sizes of extragalactic sources within $\approx$ 1\degr\ of \sgr\ can be more than an order of magnitude smaller than predicted by NE2001. 
PSRs J1746-2856 and J1746-2849 have the largest RMs measured for any object in the Milky Way beyond 0.1 pc of \sgr\ (\citealt{eatough2013}, \citealt{bower2003}, \citealt{marrone2007}); out of all known pulsars only PSR J1745-2900 has a larger RM.
All NTFs have smaller RMs, and the RMs of almost all extragalactic sources seen towards the GC are an order of magnitude smaller.

To calculate pulsar distances, NE2001 assumes a smoothly varying distribution for $n_\mathrm{e}$ which peaks at 10$\,$cm$^{-3}$.
However, there is evidence for large $n_\mathrm{e}$ variations in the GC.
Electron densities of several hundred cm$^{-3}$ have been measured with the \textit{KAO}, VLA, and the \textit{ISO} satellite (\citealt{colgan1996}, \citealt{lang2001}, \citealt{rodriguez-fernandez2001}, \citealt{cotera2005}).
Each sightline through the GC is thought to pass through at least one molecular cloud \citep{bally1988}; because molecular cloud cores\footnote{
\cite{lazio1999} argue that the ionized edges of molecular clouds could explain why scattering in the GC is patchy. They assumed that gas with a density $n\left(H_2\right) \gtrsim$ 500$\,$cm$^{-3}$ is highly ionized, however, these layers of dense gas are surrounded by more tenuous gas that could shield the inner regions from ionizing radiation (e.g., \citealt{hollenbach1999}).
The role that molecular clouds play in the scattering of background sources could be investigated with radiative transfer models that include chemical networks, e.g., $\textsc{CLOUDY}$  \citep{ferland2013}.
} are not significantly ionized, clouds leave holes in the $n_\mathrm{e}$ distribution with typical diameters of up to 20$\,$pc (FWHM; this value is based on the median diameter of molecular clouds in the CO survey by \citealt{oka2001}).
\cite{simpson2007} derived EMs along a strip at a Galactic longitude of $\approx$ 0.11\degr, and the $n_\mathrm{e}$ values they calculated vary by more than an order of magnitude.
Finally, the amount of diffuse radio emission at 10.55 GHz (fig.$\,$\ref{map.fig}) varies strongly with position, which could point to variations in the amount of free-free emission and therefore to variations in $n_\mathrm{e}$.
These results demonstrate that large $n_\mathrm{e}$ variations exist in the GC. Therefore, one may not rely on the difference between the DMs of two pulsars to estimate their relative LOS positions.

\cite{uchida1985} predicted that winding up of field lines due to differential rotation in the GC leads to an outflow perpendicular to the Galactic Plane, and produces a checkerboard pattern in RM centred on \sgr\ (fig.$\,$3 in \citealt{novak2003} and fig.$\,$14 in \citealt{law2011}). 
As discussed by Novak et al. and Law et al.\footnote{Law et al. modelled the magnetized ISM in the GC using observations of polarized diffuse radio emission. 
However, diffuse radio emission originates not only in the GC; furthermore, it can suffer from a range of depolarization effects, as discussed by, e.g., \cite{sokoloff1998}. Because of these concerns we prefer to compare our results to those presented by Uchida et al. and Novak et al.}, the signs of the RMs of NTFs agree with one of the two possible patterns.
However, the signs of the RMs of PSRs J1745-2912 and J1746-2856 contradict the particular solution selected by Uchida et al. and by Novak et al. 
The sign of RM of PSR J1746-2849 agrees with the model prediction because this pulsar lies at a slightly higher Galactic latitude than \sgr\ \citep{reid1999}, but since this latitude difference is very small, the winding up of field lines along the LOS towards PSR J1746-2849 generates only a small RM.
A large foreground RM could shift the centre of the RM pattern and disrupt the checkerboard pattern. 
However, the RMs of PSRs J1746-2856 and J1746-2849 are much larger in magnitude than the RMs of NTFs and extragalactic radio sources, which implies that the foreground contribution cannot explain the difference in sign between the RMs we observe and the RMs Uchida et al. predict.

The contribution by the Galactic foreground could dominate the RM of PSR J1745-2912, and because the RM of PSR J1745-2900 is dominated by the high $n_\mathrm{e}$ and strong magnetic field close to \sgr\ \citep{eatough2013}, we exclude these two pulsars from our analysis.
For the remaining two pulsars we calculate $\langle B_\| \rangle$, the mean LOS component of the magnetic field, weighted with the free electron density:
\begin{eqnarray}
\langle B_\| \rangle \equiv \frac{\int B_\| n_\mathrm{e}\mathrm{d}l}{\int n_\mathrm{e}\mathrm{d}l} = \frac{\mathrm{RM}}{0.81\, \mathrm{DM}}\, ,
\nonumber
\end{eqnarray}
using the definition of DM from equation~\ref{dm_definition}.
If $B_\|$ changes sign along the line of sight then locally $|B_\| |$ will be larger than $|\langle B_\| \rangle|$.
We subtract 670~cm$^{-3}$pc from the observed DMs to correct for the foreground contribution: this is the DM value predicted by NE2001 for the line of sight towards PSR J1745-2900, integrating from the Sun out to 150~pc from Sgr~A$^{\star,}$\footnote{We noticed that along this line of sight DMs do not increase monotonically with distance in NE2001.}.
The extragalactic sources that \cite{roy2008} analysed show large variations in RM, which in large part could be produced by the Galactic foreground. 
Estimating the foreground RM from these observations is very difficult, and we did not correct the pulsar RMs for the foreground contribution. 
By not correcting for the foreground RM, which is perhaps as high as 1500$\,$\radm, we overestimate $|\langle B_\| \rangle|$ by $\lesssim\, 15\%$.
For PSR J1746-2856 $\langle B_\| \rangle$ = 33$\,\mu$G, and for PSR J1746-2849 $\langle B_\| \rangle$ = 16$\,\mu$G, which are both much higher than the strength of the ordered and turbulent magnetic field in the vicinity of the Sun (1.5-2$\,\mu$G and 3-6$\,\mu$G, respectively: e.g., \citealt{haverkorn2015}).
Strong magnetic fields are not uncommon in the GC: in their analysis of the `Snake' NTF, \cite{gray1995} derived  $\langle B_\| \rangle$ $\approx$ 7$\,\mu$G for the foreground ISM.
\cite{larosa2005} analysed maps of the diffuse synchrotron emission in the GC, and showed using minimum-energy arguments that the magnetic field strength $B\, \simeq\, 6\left(k/f\right)^{2/7}\mu$G, where $k$ is the ratio of energies of cosmic ray protons and electrons, and $f$ the filling factor of the synchrotron-emitting gas. 
For reasonable values of $k$ and $f$, $B\simeq\, \left(6-80\right)\mu$G \citep{ferriere2011}.
Since the $\langle B_\| \rangle$ we derive are lower limits to $B$, our measurements rule out the weakest pervasive fields that are allowed in the analysis by LaRosa et al.
If a magnetic field with $B \sim 160\,\mu$G \citep{crocker2011} pervades the GC, including the warm ionized medium that is probed by our measurements, then the small $|\langle B_\| \rangle|$ we derive for PSR J1746-2856 implies an angle $\sim$ 12\degr\ between the magnetic field vector and the plane of the sky; towards PSR J1746-2849, which lies closer to the Galactic plane, this angle is even smaller.
If $B_\|$ changes sign along the line of sight then locally the angle between the magnetic field and the plane of the sky can be larger than the values we derived.

PSRs J1746-2849 and J1746-2856 are seen in the direction of the Radio Arc NTF (fig.$\,$\ref{map.fig}), which is known to have RMs between -5500 and -1660$\,$\radm\ (`Source A' at gl, gb = 0.16\degr, -0.13\degr; \citealt{sofue1987}) and between -200 and 1000$\,$\radm\ (`Source A$^\prime$\,' at gl, gb = 0.16\degr, -0.18\degr; \citealt{sofue1987}).
The `Plumes' are the northern and southern extension of the Radio Arc, and show RMs between -500 and 1000$\,$\radm\ (northern plume) and between -1500 and 0$\,$\radm\ (southern plume).
The RMs for the Radio Arc and the Plumes have been published by \cite{inoue1984}, \cite{sofue1987}, \cite{tsuboi1986}, \cite{yusef-zadeh1986a}, and \cite{yusef-zadeh1987}.
The large RM variations that we measured for the GC pulsars and between these pulsars and the different parts of the Radio Arc could be produced in the turbulent GC.
Variations in RM by several thousand \radm\ on scales of arcminutes have also been observed towards NTFs.(\citealt{gray1995}, \citealt{yusef-zadeh1997}, \citealt{lang1999a}, \citealt{lang1999b}). 
NTFs like the `Pelican' and the `Ripple' are known to have external Faraday screens, which implies that the observed variations in RM occur outside the NTF, in the ISM of the GC. 
Only the GC shows such large RM fluctuations, adding support that our pulsars are in the GC.
This also strengthens the case for PSR J1745-2912 lying in the GC, because its DM is very similar to the DM of PSR J1746-2856.
If PSRs J1746-2856 and J1746-2849 lie along low-scattering `corridors' through the GC this again raises the question why not more pulsars have been detected in the GC.
To answer this question requires a more detailed study of the GC environment and its pulsar population, taking into account the various instrumental effects and selection biases (building on, e.g., \citealt{chennamangalam2014}).

We cannot exclude the possibility that a foreground object in the GC adds a large RM ($\sim$ 10$^4\,$\radm) to PSRs J1746-2856 and J1746-2849. 
A helical magnetic field around the Radio Arc could be the origin of this RM, and might produce also the helical features in Stokes $I$ around the Radio Arc that were identified by \cite{yusef-zadeh1987}.
The molecular cloud G0.11-0.11 lies at gl, gb = 0.108\degr, -0.108\degr\ \citep{tsuboi1997}, in between PSRs J1746-2856 and J1746-2849.
Ablated gas from this cloud which is subsequently ionized might be an alternative explanation for the large RMs of these pulsars.
However, the inner part of G0.11-0.11 that emits in the CS $J=1\rightarrow0$ line (which traces gas with densities $n\left(H_2\right) \gtrsim$ 10$^4\,$cm$^{-3}$; \citealt{bally1988}) has a diameter of only 7 pc \citep{tsuboi1997}, while the pulsars are separated by about 29~pc on the sky if they lie in the GC.
Even though a transition region exists between the dense gas in the inner parts of G0.11-0.11 and the low-density environment of the cloud, the ionized halo surrounding G0.11-0.11 would have to be very large for it to contribute to the RMs of both pulsars.
Both scenarios require that the pulsars lie behind the Radio Arc or G0.11-0.11, which places the pulsars at the distance of \sgr\ (\citealt{ponti2010}, \citealt{roy2013}).
f correct, this provides further evidence that long ($>$~100~pc), low-scattering corridors exist in the GC.

\section{Summary}\label{summary.sec}
We studied the properties of the magnetic field in the GC by measuring the amount of Faraday rotation in the signals of four pulsars in the GC.
We observed these pulsars with the Australia Telescope Compact Array at frequencies between 4.5-6.5 GHz.
Our measurements provide more accurate positions for three of the previously known pulsars in the GC. 
The non-detection of PSR J1746-2850 implies that the flux density of this pulsar has decreased by at least a factor of thirty since its discovery by \cite{deneva2009}.
To include both the flux spectrum of each pulsar and the variation in sensitivity across the observing band we used a new maximum-likelihood based method that we will describe in an upcoming paper.
PSRs J1746-2856 and J1746-2849 have RMs$\,>\,10^4\,$rad~m$^{-2}$, the second and third largest RMs measured for any pulsar after PSR J1745-2900, and the largest RM measured for any Galactic object beyond $\sim\,0.1\,$pc of \sgr.
Their DMs, RMs, and RM variations place these pulsars within $\approx\,150\,$pc of \sgr, within the Central Molecular Zone and the footprint of the Galactic Centre Lobe.
Based on evidence from the literature we could not pinpoint the pulsar positions more accurately based on only their DMs; the smooth structure in the free electron density that is assumed in NE2001 is an oversimplification.

Differential rotation in the GC, leading to a winding-up of field lines, was predicted to produce a checkerboard pattern in the sign of RM \citep{uchida1985}; the RMs of NTFs agreed with one of the two possible solutions for this pattern (e.g., \citealt{novak2003}).
However, the signs of the RMs of PSRs J1746-2856 and J1746-2849 contradict this particular solution.
By combining the observed DMs and RMs, and correcting for the foreground contribution to DM, we derive LOS lengths of the magnetic field vector between $\sim\,16-33\,\mu$G, which is much larger than the strength of the large-scale and small-scale magnetic field in the vicinity of the Sun \citep{haverkorn2015}.
If the GC, including the warm ionized medium that is probed by our observations, is pervaded by a strong magnetic field, $B\,\sim\,160\,\mu$G \citep{crocker2011}, then our measurements imply that the magnetic field makes an angle $\lesssim\,12$\degr\ with the plane of the sky.
If the direction of the LOS magnetic field component changes sign along the line of sight, this inclination angle will be larger.
Large changes in RM on scales of arcminutes have been observed previously towards NTFs; since the GC is the only location where such large RM variations are known to exist, this provides additional evidence that the target pulsars must lie inside the GC. 
The large RMs of PSRs J1746-2856 and J1746-2849 could be produced by foreground objects such as the Radio Arc or an ablated, ionized halo that surrounds the molecular cloud G0.11-0.11.
This implies that these two pulsars must lie close to or even beyond \sgr, more than 100 pc into the GC, which would be additional evidence for the existence of long, low-scattering corridors through the GC.
If this is indeed the case, then future, sensitive observations should be able to detect more pulsars in the GC (e.g., \citealt{chennamangalam2014}, \citealt{eatough2015}).

\section*{Acknowledgements}
We would like to thank the staff at CSIRO Astronomy and Space Science, and in particular Robin Wark, Jamie Stevens, and Phil Edwards, for their support of this project. 
Fig.$\,$\ref{map.fig} was prepared using a script provided by Bernd Klein (Max Planck Institute for Radio Astronomy).
The Australia Telescope Compact Array is part of the Australia Telescope National Facility which is funded by the Commonwealth of Australia for operation as a National Facility managed by CSIRO.


\begin{thebibliography}{}

\bibitem[\protect\citeauthoryear{{Avni}}{{Avni}}{1976}]{avni1976}
{Avni} Y.,  1976, \apj, 210, 642

\bibitem[\protect\citeauthoryear{{Bally}, {Stark}, {Wilson} \&
  {Henkel}}{{Bally} et~al.}{1988}]{bally1988}
{Bally} J.,  {Stark} A.~A.,  {Wilson} R.~W.,    {Henkel} C.,  1988, \apj, 324,
  223

\bibitem[\protect\citeauthoryear{{Bates}, {Lorimer} \& {Verbiest}}{{Bates}
  et~al.}{2013}]{Bates2013}
{Bates} S.~D.,  {Lorimer} D.~R.,    {Verbiest} J.~P.~W.,  2013, \mnras, 431,
  1352

\bibitem[\protect\citeauthoryear{{Bower}, {Backer} \& {Sramek}}{{Bower}
  et~al.}{2001}]{bower2001}
{Bower} G.~C.,  {Backer} D.~C.,    {Sramek} R.~A.,  2001, \apj, 558, 127

\bibitem[\protect\citeauthoryear{{Bower}, {Wright}, {Falcke} \&
  {Backer}}{{Bower} et~al.}{2003}]{bower2003}
{Bower} G.~C.,  {Wright} M.~C.~H.,  {Falcke} H.,    {Backer} D.~C.,  2003,
  \apj, 588, 331

\bibitem[\protect\citeauthoryear{{Brentjens} \& {de Bruyn}}{{Brentjens} \& {de
  Bruyn}}{2005}]{brentjens2005}
{Brentjens} M.~A.,  {de Bruyn} A.~G.,  2005, \aap, 441, 1217

\bibitem[\protect\citeauthoryear{{Chennamangalam} \&
  {Lorimer}}{{Chennamangalam} \& {Lorimer}}{2014}]{chennamangalam2014}
{Chennamangalam} J.,  {Lorimer} D.~R.,  2014, \mnras, 440, L86

\bibitem[\protect\citeauthoryear{{Colgan}, {Erickson}, {Simpson}, {Haas} \&
  {Morris}}{{Colgan} et~al.}{1996}]{colgan1996}
{Colgan} S.~W.~J.,  {Erickson} E.~F.,  {Simpson} J.~P.,  {Haas} M.~R.,
  {Morris} M.,  1996, \apj, 470, 882

\bibitem[\protect\citeauthoryear{{Cordes} \& {Lazio}}{{Cordes} \&
  {Lazio}}{2002}]{cordes2002}
{Cordes} J.~M.,  {Lazio} T.~J.~W.,  2002, ArXiv:0207156

\bibitem[\protect\citeauthoryear{{Cotera}, {Colgan}, {Simpson} \&
  {Rubin}}{{Cotera} et~al.}{2005}]{cotera2005}
{Cotera} A.~S.,  {Colgan} S.~W.~J.,  {Simpson} J.~P.,    {Rubin} R.~H.,  2005,
  \apj, 622, 333

\bibitem[\protect\citeauthoryear{{Crocker}, {Jones}, {Aharonian}, {Law},
  {Melia}, {Oka} \& {Ott}}{{Crocker} et~al.}{2011}]{crocker2011}
{Crocker} R.~M.,  {Jones} D.~I.,  {Aharonian} F.,  {Law} C.~J.,  {Melia} F.,
  {Oka} T.,    {Ott} J.,  2011, \mnras, 413, 763

\bibitem[\protect\citeauthoryear{{Deneva}, {Cordes} \& {Lazio}}{{Deneva}
  et~al.}{2009}]{deneva2009}
{Deneva} J.~S.,  {Cordes} J.~M.,    {Lazio} T.~J.~W.,  2009, \apjl, 702, L177

\bibitem[\protect\citeauthoryear{{Eatough}, {Lazio}, {Casanellas},
  {Chatterjee}, {Cordes}, {Demorest}, {Kramer}, {Lee}, {Liu}, {Ransom} \&
  {Wex}}{{Eatough} et~al.}{2015}]{eatough2015}
{Eatough} R.,  {Lazio} T.~J.~W.,  {Casanellas} J.,  {Chatterjee} S.,  {Cordes}
  J.~M.,  {Demorest} P.~B.,  {Kramer} M.,  {Lee} K.~J.,  {Liu} K.,  {Ransom}
  S.~M.,    {Wex} N.,  2015, Advancing Astrophysics with the Square Kilometre
  Array (AASKA14), p.~45

\bibitem[\protect\citeauthoryear{{Eatough}, {Falcke}, {Karuppusamy}, {Lee},
  {Champion}, {Keane}, {Desvignes}, {Schnitzeler}, {Spitler}, {Kramer},
  {Klein}, {Bassa}, {Bower} \& {Brunthaler}}{{Eatough}
  et~al.}{2013}]{eatough2013}
{Eatough} R.~P.,  {Falcke} H.,  {Karuppusamy} R.,  {Lee} K.~J.,  {Champion}
  D.~J.,  {Keane} E.~F.,  {Desvignes} G.,  {Schnitzeler} D.~H.~F.~M.,
  {Spitler} L.~G.,  {Kramer} M.,  {Klein} B.,  {Bassa} C.,  {Bower} G.~C.,
  {Brunthaler} 2013, \nat, 501, 391

\bibitem[\protect\citeauthoryear{{Ferland}, {Porter}, {van Hoof}, {Williams},
  {Abel}, {Lykins}, {Shaw}, {Henney} \& {Stancil}}{{Ferland}
  et~al.}{2013}]{ferland2013}
{Ferland} G.~J.,  {Porter} R.~L.,  {van Hoof} P.~A.~M.,  {Williams} R.~J.~R.,
  {Abel} N.~P.,  {Lykins} M.~L.,  {Shaw} G.,  {Henney} W.~J.,    {Stancil}
  P.~C.,  2013, \rmxaa, 49, 137

\bibitem[\protect\citeauthoryear{{Ferri{\`e}re}}{{Ferri{\`e}re}}{2011}]{ferriere2011}
{Ferri{\`e}re} K.,  2011, in {Morris} M.~R.,  {Wang} Q.~D.,   {Yuan} F.,  eds,
  The Galactic Center: a Window to the Nuclear Environment of Disk Galaxies
  Vol.~439 of Astronomical Society of the Pacific Conference Series, {Magnetic
  Fields in the Galactic Center}.
p.~39

\bibitem[\protect\citeauthoryear{{Gray}, {Nicholls}, {Ekers} \& {Cram}}{{Gray}
  et~al.}{1995}]{gray1995}
{Gray} A.~D.,  {Nicholls} J.,  {Ekers} R.~D.,    {Cram} L.~E.,  1995, \apj,
  448, 164

\bibitem[\protect\citeauthoryear{{Haverkorn}}{{Haverkorn}}{2015}]{haverkorn2015}
{Haverkorn} M.,  2015, in {Lazarian} A.,  {de Gouveia Dal Pino} E.~M.,
  {Melioli} C.,  eds, Astrophysics and Space Science Library Vol.~407 of
  Astrophysics and Space Science Library, {Magnetic Fields in the Milky Way}.
p.~483

\bibitem[\protect\citeauthoryear{{Hollenbach} \& {Tielens}}{{Hollenbach} \&
  {Tielens}}{1999}]{hollenbach1999}
{Hollenbach} D.~J.,  {Tielens} A.~G.~G.~M.,  1999, Reviews of Modern Physics,
  71, 173

\bibitem[\protect\citeauthoryear{{Inoue}, {Takahashi}, {Tabara}, {Kato} \&
  {Tsuboi}}{{Inoue} et~al.}{1984}]{inoue1984}
{Inoue} M.,  {Takahashi} T.,  {Tabara} H.,  {Kato} T.,    {Tsuboi} M.,  1984,
  \pasj, 36, 633

\bibitem[\protect\citeauthoryear{{Johnston}, {Kramer}, {Lorimer}, {Lyne},
  {McLaughlin}, {Klein} \& {Manchester}}{{Johnston}
  et~al.}{2006}]{johnston2006}
{Johnston} S.,  {Kramer} M.,  {Lorimer} D.~R.,  {Lyne} A.~G.,  {McLaughlin} M.,
   {Klein} B.,    {Manchester} R.~N.,  2006, \mnras, 373, L6

\bibitem[\protect\citeauthoryear{{Kennea}, {Burrows}, {Kouveliotou}, {Palmer},
  {G{\"o}{\u g}{\"u}{\c s}}, {Kaneko}, {Evans}, {Degenaar}, {Reynolds},
  {Miller}, {Wijnands}, {Mori} \& {Gehrels}}{{Kennea}
  et~al.}{2013}]{kennea2013}
{Kennea} J.~A.,  {Burrows} D.~N.,  {Kouveliotou} C.,  {Palmer} D.~M.,
  {G{\"o}{\u g}{\"u}{\c s}} E.,  {Kaneko} Y.,  {Evans} P.~A.,  {Degenaar} N.,
  {Reynolds} M.~T.,  {Miller} J.~M.,  {Wijnands} R.,  {Mori} K.,    {Gehrels}
  N.,  2013, \apjl, 770, L24

\bibitem[\protect\citeauthoryear{{Lang}, {Anantharamaiah}, {Kassim} \&
  {Lazio}}{{Lang} et~al.}{1999}]{lang1999b}
{Lang} C.~C.,  {Anantharamaiah} K.~R.,  {Kassim} N.~E.,    {Lazio} T.~J.~W.,
  1999, \apjl, 521, L41

\bibitem[\protect\citeauthoryear{{Lang}, {Goss} \& {Morris}}{{Lang}
  et~al.}{2001}]{lang2001}
{Lang} C.~C.,  {Goss} W.~M.,    {Morris} M.,  2001, \aj, 121, 2681

\bibitem[\protect\citeauthoryear{{Lang}, {Morris} \& {Echevarria}}{{Lang}
  et~al.}{1999}]{lang1999a}
{Lang} C.~C.,  {Morris} M.,    {Echevarria} L.,  1999, \apj, 526, 727

\bibitem[\protect\citeauthoryear{{LaRosa}, {Brogan}, {Shore}, {Lazio}, {Kassim}
  \& {Nord}}{{LaRosa} et~al.}{2005}]{larosa2005}
{LaRosa} T.~N.,  {Brogan} C.~L.,  {Shore} S.~N.,  {Lazio} T.~J.,  {Kassim}
  N.~E.,    {Nord} M.~E.,  2005, \apjl, 626, L23

\bibitem[\protect\citeauthoryear{{Law}, {Brentjens} \& {Novak}}{{Law}
  et~al.}{2011}]{law2011}
{Law} C.~J.,  {Brentjens} M.~A.,    {Novak} G.,  2011, \apj, 731, 36

\bibitem[\protect\citeauthoryear{{Lazaridis}, {Jessner}, {Kramer}, {Stappers},
  {Lyne}, {Jordan}, {Serylak} \& {Zensus}}{{Lazaridis}
  et~al.}{2008}]{lazaridis2008}
{Lazaridis} K.,  {Jessner} A.,  {Kramer} M.,  {Stappers} B.~W.,  {Lyne} A.~G.,
  {Jordan} C.~A.,  {Serylak} M.,    {Zensus} J.~A.,  2008, \mnras, 390, 839

\bibitem[\protect\citeauthoryear{{Lazio}, {Anantharamaiah}, {Goss}, {Kassim} \&
  {Cordes}}{{Lazio} et~al.}{1999}]{lazio1999}
{Lazio} T.~J.~W.,  {Anantharamaiah} K.~R.,  {Goss} W.~M.,  {Kassim} N.~E.,
  {Cordes} J.~M.,  1999, \apj, 515, 196

\bibitem[\protect\citeauthoryear{{Lorimer}, {Yates}, {Lyne} \&
  {Gould}}{{Lorimer} et~al.}{1995}]{Lorimer1995}
{Lorimer} D.~R.,  {Yates} J.~A.,  {Lyne} A.~G.,    {Gould} D.~M.,  1995,
  \mnras, 273, 411

\bibitem[\protect\citeauthoryear{{Marrone}, {Moran}, {Zhao} \& {Rao}}{{Marrone}
  et~al.}{2007}]{marrone2007}
{Marrone} D.~P.,  {Moran} J.~M.,  {Zhao} J.-H.,    {Rao} R.,  2007, \apjl, 654,
  L57

\bibitem[\protect\citeauthoryear{{Mori}, {Gotthelf}, {Zhang}, {An}, {Baganoff},
  {Barri{\`e}re}, {Beloborodov}, {Boggs}, {Christensen}, {Craig}, {Dufour},
  {Grefenstette} \& {Hailey} C.~J.}{{Mori} et~al.}{2013}]{mori2013}
{Mori} K.,  {Gotthelf} E.~V.,  {Zhang} S.,  {An} H.,  {Baganoff} F.~K.,
  {Barri{\`e}re} N.~M.,  {Beloborodov} A.~M.,  {Boggs} S.~E.,  {Christensen}
  F.~E.,  {Craig} W.~W.,  {Dufour} F.,  {Grefenstette} B.~W.,    {Hailey} C.~J.
  e.~a.,  2013, \apjl, 770, L23

\bibitem[\protect\citeauthoryear{{Noutsos}, {Karastergiou}, {Kramer},
  {Johnston} \& {Stappers}}{{Noutsos} et~al.}{2009}]{noutsos2009}
{Noutsos} A.,  {Karastergiou} A.,  {Kramer} M.,  {Johnston} S.,    {Stappers}
  B.~W.,  2009, \mnras, 396, 1559

\bibitem[\protect\citeauthoryear{{Novak}, {Chuss}, {Renbarger}, {Griffin},
  {Newcomb}, {Peterson}, {Loewenstein}, {Pernic} \& {Dotson}}{{Novak}
  et~al.}{2003}]{novak2003}
{Novak} G.,  {Chuss} D.~T.,  {Renbarger} T.,  {Griffin} G.~S.,  {Newcomb}
  M.~G.,  {Peterson} J.~B.,  {Loewenstein} R.~F.,  {Pernic} D.,    {Dotson}
  J.~L.,  2003, \apjl, 583, L83

\bibitem[\protect\citeauthoryear{{Oka}, {Hasegawa}, {Sato}, {Tsuboi},
  {Miyazaki} \& {Sugimoto}}{{Oka} et~al.}{2001}]{oka2001}
{Oka} T.,  {Hasegawa} T.,  {Sato} F.,  {Tsuboi} M.,  {Miyazaki} A.,
  {Sugimoto} M.,  2001, \apj, 562, 348

\bibitem[\protect\citeauthoryear{{Ponti}, {Terrier}, {Goldwurm}, {Belanger} \&
  {Trap}}{{Ponti} et~al.}{2010}]{ponti2010}
{Ponti} G.,  {Terrier} R.,  {Goldwurm} A.,  {Belanger} G.,    {Trap} G.,  2010,
  \apj, 714, 732

\bibitem[\protect\citeauthoryear{{Reid}, {Menten}, {Brunthaler}, {Zheng},
  {Dame}, {Xu}, {Wu}, {Zhang}, {Sanna}, {Sato}, {Hachisuka}, {Choi}, {Immer},
  {Moscadelli}, {Rygl} \& {Bartkiewicz}}{{Reid} et~al.}{2014}]{reid2014}
{Reid} M.~J.,  {Menten} K.~M.,  {Brunthaler} A.,  {Zheng} X.~W.,  {Dame} T.~M.,
   {Xu} Y.,  {Wu} Y.,  {Zhang} B.,  {Sanna} A.,  {Sato} M.,  {Hachisuka} K.,
  {Choi} Y.~K.,  {Immer} K.,  {Moscadelli} L.,  {Rygl} K.~L.~J.,
  {Bartkiewicz} A.,  2014, \apj, 783, 130

\bibitem[\protect\citeauthoryear{{Reid}, {Readhead}, {Vermeulen} \&
  {Treuhaft}}{{Reid} et~al.}{1999}]{reid1999}
{Reid} M.~J.,  {Readhead} A.~C.~S.,  {Vermeulen} R.~C.,    {Treuhaft} R.~N.,
  1999, \apj, 524, 816

\bibitem[\protect\citeauthoryear{{Rodr{\'{\i}}guez-Fern{\'a}ndez},
  {Mart{\'{\i}}n-Pintado} \& {de Vicente}}{{Rodr{\'{\i}}guez-Fern{\'a}ndez}
  et~al.}{2001}]{rodriguez-fernandez2001}
{Rodr{\'{\i}}guez-Fern{\'a}ndez} N.~J.,  {Mart{\'{\i}}n-Pintado} J.,    {de
  Vicente} P.,  2001, \aap, 377, 631

\bibitem[\protect\citeauthoryear{{Roy}}{{Roy}}{2013}]{roy2013}
{Roy} S.,  2013, \apj, 773, 67

\bibitem[\protect\citeauthoryear{{Roy}, {Pramesh Rao} \& {Subrahmanyan}}{{Roy}
  et~al.}{2008}]{roy2008}
{Roy} S.,  {Pramesh Rao} A.,    {Subrahmanyan} R.,  2008, \aap, 478, 435

\bibitem[\protect\citeauthoryear{{Sault}, {Teuben} \& {Wright}}{{Sault}
  et~al.}{1995}]{sault1995}
{Sault} R.~J.,  {Teuben} P.~J.,    {Wright} M.~C.~H.,  1995, in {Shaw} R.~A.,
  {Payne} H.~E.,   {Hayes} J.~J.~E.,  eds, Astronomical Data Analysis Software
  and Systems IV Vol.~77 of Astronomical Society of the Pacific Conference
  Series, {A Retrospective View of MIRIAD}.
p.~433

\bibitem[\protect\citeauthoryear{{Schnitzeler}}{{Schnitzeler}}{2012}]{schnitzeler2012}
{Schnitzeler} D.~H.~F.~M.,  2012, \mnras, 427, 664

\bibitem[\protect\citeauthoryear{{Schnitzeler} \& {Lee}}{{Schnitzeler} \&
  {Lee}}{2015}]{schnitzeler2015}
{Schnitzeler} D.~H.~F.~M.,  {Lee} K.~J.,  2015, \mnras, 447, L26

\bibitem[\protect\citeauthoryear{{Seiradakis}, {Reich}, {Wielebinski},
  {Lasenby} \& {Yusef-Zadeh}}{{Seiradakis} et~al.}{1989}]{seiradakis1989}
{Seiradakis} J.~H.,  {Reich} W.,  {Wielebinski} R.,  {Lasenby} A.~N.,
  {Yusef-Zadeh} F.,  1989, \aaps, 81, 291

\bibitem[\protect\citeauthoryear{{Shannon} \& {Johnston}}{{Shannon} \&
  {Johnston}}{2013}]{shannon2013}
{Shannon} R.~M.,  {Johnston} S.,  2013, \mnras, 435, L29

\bibitem[\protect\citeauthoryear{{Simpson}, {Colgan}, {Cotera}, {Erickson},
  {Hollenbach}, {Kaufman} \& {Rubin}}{{Simpson} et~al.}{2007}]{simpson2007}
{Simpson} J.~P.,  {Colgan} S.~W.~J.,  {Cotera} A.~S.,  {Erickson} E.~F.,
  {Hollenbach} D.~J.,  {Kaufman} M.~J.,    {Rubin} R.~H.,  2007, \apj, 670,
  1115

\bibitem[\protect\citeauthoryear{{Sofue} \& {Handa}}{{Sofue} \&
  {Handa}}{1984}]{sofue1984}
{Sofue} Y.,  {Handa} T.,  1984, \nat, 310, 568

\bibitem[\protect\citeauthoryear{{Sofue}, {Reich}, {Inoue} \&
  {Seiradakis}}{{Sofue} et~al.}{1987}]{sofue1987}
{Sofue} Y.,  {Reich} W.,  {Inoue} M.,    {Seiradakis} J.~H.,  1987, \pasj, 39,
  95

\bibitem[\protect\citeauthoryear{{Sokoloff}, {Bykov}, {Shukurov},
  {Berkhuijsen}, {Beck} \& {Poezd}}{{Sokoloff} et~al.}{1998}]{sokoloff1998}
{Sokoloff} D.~D.,  {Bykov} A.~A.,  {Shukurov} A.,  {Berkhuijsen} E.~M.,  {Beck}
  R.,    {Poezd} A.~D.,  1998, \mnras, 299, 189

\bibitem[\protect\citeauthoryear{{Tsuboi}, {Inoue}, {Handa}, {Tabara}, {Kato},
  {Sofue} \& {Kaifu}}{{Tsuboi} et~al.}{1986}]{tsuboi1986}
{Tsuboi} M.,  {Inoue} M.,  {Handa} T.,  {Tabara} H.,  {Kato} T.,  {Sofue} Y.,
   {Kaifu} N.,  1986, \aj, 92, 818

\bibitem[\protect\citeauthoryear{{Tsuboi}, {Ukita} \& {Handa}}{{Tsuboi}
  et~al.}{1997}]{tsuboi1997}
{Tsuboi} M.,  {Ukita} N.,    {Handa} T.,  1997, \apj, 481, 263

\bibitem[\protect\citeauthoryear{{Uchida}, {Sofue} \& {Shibata}}{{Uchida}
  et~al.}{1985}]{uchida1985}
{Uchida} Y.,  {Sofue} Y.,    {Shibata} K.,  1985, \nat, 317, 699

\bibitem[\protect\citeauthoryear{{Wilks}}{{Wilks}}{1938}]{wilks1938}
{Wilks} S.~S.,  1938, Ann. Math. Stat., 9, 60

\bibitem[\protect\citeauthoryear{{Yusef-Zadeh} \& {Morris}}{{Yusef-Zadeh} \&
  {Morris}}{1987}]{yusef-zadeh1987}
{Yusef-Zadeh} F.,  {Morris} M.,  1987, \apj, 322, 721

\bibitem[\protect\citeauthoryear{{Yusef-Zadeh}, {Morris} \&
  {Chance}}{{Yusef-Zadeh} et~al.}{1984}]{yusef-zadeh1984}
{Yusef-Zadeh} F.,  {Morris} M.,    {Chance} D.,  1984, \nat, 310, 557

\bibitem[\protect\citeauthoryear{{Yusef-Zadeh}, {Morris}, {Slee} \&
  {Nelson}}{{Yusef-Zadeh} et~al.}{1986}]{yusef-zadeh1986a}
{Yusef-Zadeh} F.,  {Morris} M.,  {Slee} O.~B.,    {Nelson} G.~J.,  1986, \apj,
  310, 689

\bibitem[\protect\citeauthoryear{{Yusef-Zadeh}, {Wardle} \&
  {Parastaran}}{{Yusef-Zadeh} et~al.}{1997}]{yusef-zadeh1997}
{Yusef-Zadeh} F.,  {Wardle} M.,    {Parastaran} P.,  1997, \apjl, 475, L119

\end{thebibliography}
\end{document}